# Measuring the magnetization of three monolayer thick Co islands and films by X-ray dichroism

A. Mascaraque,[1] L. Aballe,[2] J. F. Marco,[3] T. O. Menteş,[2] F. El Gabaly,[4] C. Klein,[5] A. K. Schmid,[5] K. F. McCarty,[4] A. Locatelli,[2] and J. de la Figuera[3, 6]

[1]Universidad Complutense de Madrid, Madrid 28040, Spain

[2]Elettra Sincrotrone S.C.p.A, Trieste, Italy

[3]Instituto de Química-Física "Rocasolano", CSIC, Madrid 28006, Spain

[4]Sandia National Laboratories, Livermore, California 94550, USA

[5]Lawrence Berkeley National Laboratory, Berkeley, California 94720, USA

[6]Centro de Microanálisis de Materiales, Universidad Autónoma de Madrid, Madrid 28049, Spain

**Abstract**

Co islands and films are characterized by x-ray magnetic circular dichroism photoemission electron microscopy (XMCD-PEEM). The spatial resolution capabilties of the technique together with atomic growth control permit obtaining perfectly flat triangular islands with a given thickness (3 ML), very close to an abrupt spin-reorientation transition. The magnetic domain configurations are found to depend on island size: while small islands can be magnetized in a single-domain state, larger islands show more complex patterns. Furthermore, the magnetization pattern of the larger islands presents a common chirality. By means of dichroic spectro-microscopy at the Co L absorption edges, an experimental estimate of the ratio of the spin- and orbital magnetic moment for three monolayer thick films is obtained.

Understanding the growth of magnetic nanostructures with well-controlled properties is crucial for the development of spintronics and to enable decreasing bit size in magnetic storage media [1]. In particular, being able to engineer anisotropy by controlling growth processes at the atomic scale is one important goal [2]. However, roughness, as well as size and shape dispersion of as-grown structures complicates understanding the interplay of morphology and other properties, and adversely impacts the reproducibility of fabrication processes. Ultimately, these problems limit potential uses in actual devices. This work is performed on nanostructures of well controlled thickness, as close as possible to a spin reorientation transition.

Co/Ru(0001) has been the subject of many studies, ranging from multilayers to nanoparticles [3-14]. In spite of this focussed interest, only recently the presence of abrupt spin reorientation transitions in ultrathin Co/Ru(0001) films has been reported [13,14]: while in one atomic monolayer (ML) thick films the orientation of the remanent magnetization is parallel to the surface (i.e., the magnetic easy axis is in the film plane), films two layers thick show an out-of-plane magnetization (perpendicular magnetic anisotropy, PMA [15]). Films thicker than two layers present again an in-plane magnetization. The fact that spin reorientation transitions can be used to tune the magnetic properties of films and microscopic structures through precise control of the layer thickness is one of the driving motivations for looking deeper into the magnetic properties of the Co/Ru system close to this previously unobserved transition. For this purpose, both the value of the magnetic anisotropy energy as a function of atomic layer thickness and the atomic magnetic moment in the structures are important. Measurements of the atomic magnetic moment in microscopic structures with precisely controlled thickness in the range of a few atomic monolayers are rare. More commonly, averaging techniques are employed to determine the value of the magnetic moment with high accuracy, at the expense of higher uncertainty regarding the detailed morphology of the samples [11,12]. In the work presented here, micron-sized, triangular, three monolayer-thick islands and films with in-plane magnetization are prepared [13,16]. Analyzing individual microscopic Co islands, we observe the effect of island size on magnetic domain configurations, and measure the ratio of the orbital and spin components of the magnetic moment.

Cobalt on ruthenium is a typical Stranski-Krastanov system [5,8], with a well-known tendency to form three-dimensional Co islands separated by a wetting layer. In order to investigate films and islands of precisely controlled uniform thickness, the formation of 3D islands must be suppressed. As it was shown recently [16], Co layer-by-layer growth up to 10 ML can be achieved on micron sized substrate terraces by carefully tuning the growth conditions. Using the sample preparation methods described in Ref. [16], we focus in this study on the 3 ML case, the thinnest film with in-plane magnetization and Curie temperature above room temperature (RT). Films are grown on suitably large Ru terraces, choosing an appropriate substrate temperature and Co flux, and stopping the evaporation when the desired morphology of 3 ML triangles on 2 ML is reached. Controlling the coverage provides islands ranging in size from a few hundred nanometers up to above 1 μm, as shown in Fig. 1.

The experiments were done at the spectroscopic photoemission and low energy electron microscope system [17] at the Nanospectroscopy beamline of the storage ring Elettra. The measurements are performed in two consecutive steps. First, illuminating the sample surface with an electron beam, the growth is monitored in real time by low-energy electron microscopy (LEEM). Co is evaporated in ultra-high vacuum ( base pressure below $2 \times 10^{-10}$ torr) using an e-beam heated doser at a rate of ≈ 0.25 ML/min, while the substrate is kept at ≈ 470 K [16]. The island thickness and size is controlled by the Co flux, substrate temperature, and total Co dose.

We determine the domain patterns on the cobalt islands by means of X-ray magnetic circular dichroism with photoemission electron microscopy (XMCD-PEEM [18,19]), one of the few techniques for imaging *in-situ* surface ferromagnetic domain patterns. The use of XMCD-PEEM is mandatory for this experiment, since Co/Ru(0001) films do not grow layer-by-layer except in very particular substrate areas with extremely low step density. Thus any experiment performed on Co/Ru films applying traditional experimental dichroic techniques cannot be used to extract meaningful physical data at the ultra-thin film limit.

XMCD microscopy was performed at room temperature, by illuminating with X-rays at the Co $L_3$-edge energy the region previously observed by LEEM, and forming an image with the photoexcited secondary electrons (i.e., a partial yield X-ray absorption image). In the instrument used, the X-ray beam incidence is fixed at 74º with respect to the surface normal, and thus our experiment is mostly sensitive to the in-plane component of the sample magnetization. In order to obtain magnetic contrast, pairs of images with circularly polarized X-rays of opposite polarization are acquired. In order to suppress topographic features and enhance magnetic contrast, pixel by pixel difference images are obtained [see Fig. 1(b,d)]. Surface regions with a component of the magnetization parallel (antiparallel) to the beam direction result in bright (dark) areas in the images.

Fig. 1(a,c) shows two sets of 3 ML thick triangular islands. In both cases, two substrate terraces separated by a monoatomic step are seen. In each terrace, triangular islands dominantly point in the same direction, as a result of their common fcc stacking sequence [16]. A few exceptions (such as the island marked with a circle) are ordered in an hcp stacking sequence and point in the opposite direction.

Magnetization patterns measured in remanence on as-grown 3ML thick islands are shown in Fig 1b,d. There is a clear correlation of magnetization and island size: smaller islands tend to present a uniform magnetic contrast, and to have the same contrast as neighbour islands [see Fig. 1(b)]. This suggests that small islands are in single domain states, and that closely spaced islands are magnetically coupled, sometimes in groups including several islands. On the other hand, as-grown larger islands [with sides larger than one micrometer, see Fig. 1(d)] often present a non-uniform magnetic contrast, indicating multi-domain magnetization states[20].

Magnetization patterns of the same area were also measured after applying a 0.1 T magnetic field pulse. Such field modifies the magnetization but is not enough to achieve a uniform magnetization of the larger islands (Fig. 2(a)). It is worth remarking that the pattern in all the islands presents the same chirality. Although further characterization is needed to fully understand these patterns, we tentatively attribute the common chirality to the interplay of the island shape and orientation with the magnetic field direction, as revealed by micromagnetic simulations [20]. We predict that this effect should be general, and thus be present for any well-defined triangular island. The reason why it has not been observed so far in triangular islands ([21,22]) might be related to the need of very perfect islands, difficult to achieve with common nanostructuring techniques such as focused ion beam (FIB). The patterns change with time: several consecutive XMCD-PEEM images of the same islands [see Fig.2(b-d)] show the evolution of the patterns on several islands.

The atomic magnetic moment of nanostructures often depends on their size and/or morphology. This motivates ongoing experimental efforts to determine the magnetic moment of well-defined systems of reduced dimensions, where comparison with first principles calculations can be performed. In this context, a unique advantage of XMCD-PEEM over other magnetic microscopic techniques is the ability to probe the magnetic moment *in-situ* [18].

Co L-edge absorption spectra from small islands were obtained by acquiring a sequence of images at different photon energies and integrating the intensity from individual islands. Experimental conditions, including slow drifting of the positions of the X-ray beam and the sample manipulator, place an upper limit in the acquisition time.. Given the limited integration area (to match island size or domain size in larger islands) and time (limited by position drift), the signal to noise ratio does not permit reliable measurements of the dichroic difference on individual 3 ML islands, that has thus been measured on continuous 3 ML thick films. Careful in-situ LEEM observation is crucial in order to achieve the growth of high-quality 3 ML films, since they are metastable towards formation of 3D islands. The growth of extended, continuous, 3 ML films is limited to rare substrate regions where micron-sized atomically flat terraces exist. In most of the surface, the higher substrate step density enhances the nucleation and growth of the more stable 3D islands [23].

A LEEM image of a 3 ML Co film can be seen on Fig. 3(a). In the upper-right corner, narrower terraces have resulted in 3D growth, while two large terraces in the rest of the image show only a few small 4 ML islands. On large terraces, over 95% of the surface is covered with 3 ML. The film has fcc structure, except in small regions (less than 300 nm wide) adjacent to substrate steps, where it contains stacking faults as described in Ref. [16]. After Co deposition, an external magnetic field pulse of 0.1 T is applied in the direction of the X-ray beam [24].

The preparation presents large regions with homogeneous thickness, structure, and magnetization, permitting the use of larger integration windows, and thus achieving reasonable signal to noise ratio with acceptably short integration times. XMCD spectra of epitaxial 3 ML Co/Ru(0001), for both circular light polarizations, are shown in Fig. 3(b). It is possible to extract experimental estimates of the values of the orbital and the spin part of the Co magnetic moment, $L$ and $S$, from XMCD data. The absolute values of the orbital and the spin part of the Co magnetic moment are not determined, since measuring in remanence the sample might not be completely saturated [25]. Applying the sum rules [26-28] to the dichroic difference spectrum Fig. 3(c) leads to $L/S = 0.08 \pm 0.02$, where we have disregarded effects from the magnetic dipole operator [28,29]. We note that this value is affected by details of the beam motion correction applied to the XAS spectra [30]: changing the beam motion correction within reasonable limits gives values for $L/S$ within the 0.06-0.19 bracket, skewed towards the low value for a flat background after the L absorption edge. Despite the technical limitations encountered, this work demonstrates that the orbital and spin magnetic moments of ultrathin films and nanostructures can be obtained by XMCD-PEEM, and hopes to open a new research path for other nanostructured systems.

Due to the absence of other experimentally determined magnetic moments for 3 ML fcc-Co/Ru(0001) films, our results can only be compared to first-principles calculations. A fully relativistic calculation of the Co moments in this structure, based on the screened Korringa-Kohn-Rostoker (SKKR) method, gives magnetic moments of respectively 1.62 and 0.10 $\mu_B$ for the average spin and orbital component [31]. This theoretical value agrees very well with our experimental result. Our experimental result also brackets the experimentally determined ratio of 0.095 for the magnetic moments in bulk hcp Co [28] and agrees with fcc Co [32], although this last measurement was obtained from a multilayered sample.

In summary, we have observed novel effects in the magnetization of triangular flat islands of Co on Ru, by imaging magnetic domains with XMCD PEEM at the Co $L_3$-edge energy. While small islands present uniform magnetization, larger islands (more than 1 μm in size) have more complex domain patterns. After applying a magnetic field, the magnetization patterns of larger islands show a common chirality, attributed to the interplay of their triangular shape and orientation with the external field direction. Furthermore, dichroic spectra have been obtained by XMCD-PEEM spectromicroscopy, and applying the sum rules the ratio of the orbital and spin part of the magnetic moment in the 3 ML film has been estimated to be 0.08, in good agreement with theoretical calculations.

**Acknowledgments**

This research was partly supported by the U. S. Department of Energy under contract No. DE-AC04-94AL85000 and DE-AC02—05CH11231, by the Spanish Ministry of Education and Science under Project No. MAT2006-13149-C02-02, and by the Comunidad Autónoma de Madrid and the CSIC under project No. CCG07-CSIC-MAT-2030 and project No. S-0505/MAT/0194.

# Figures

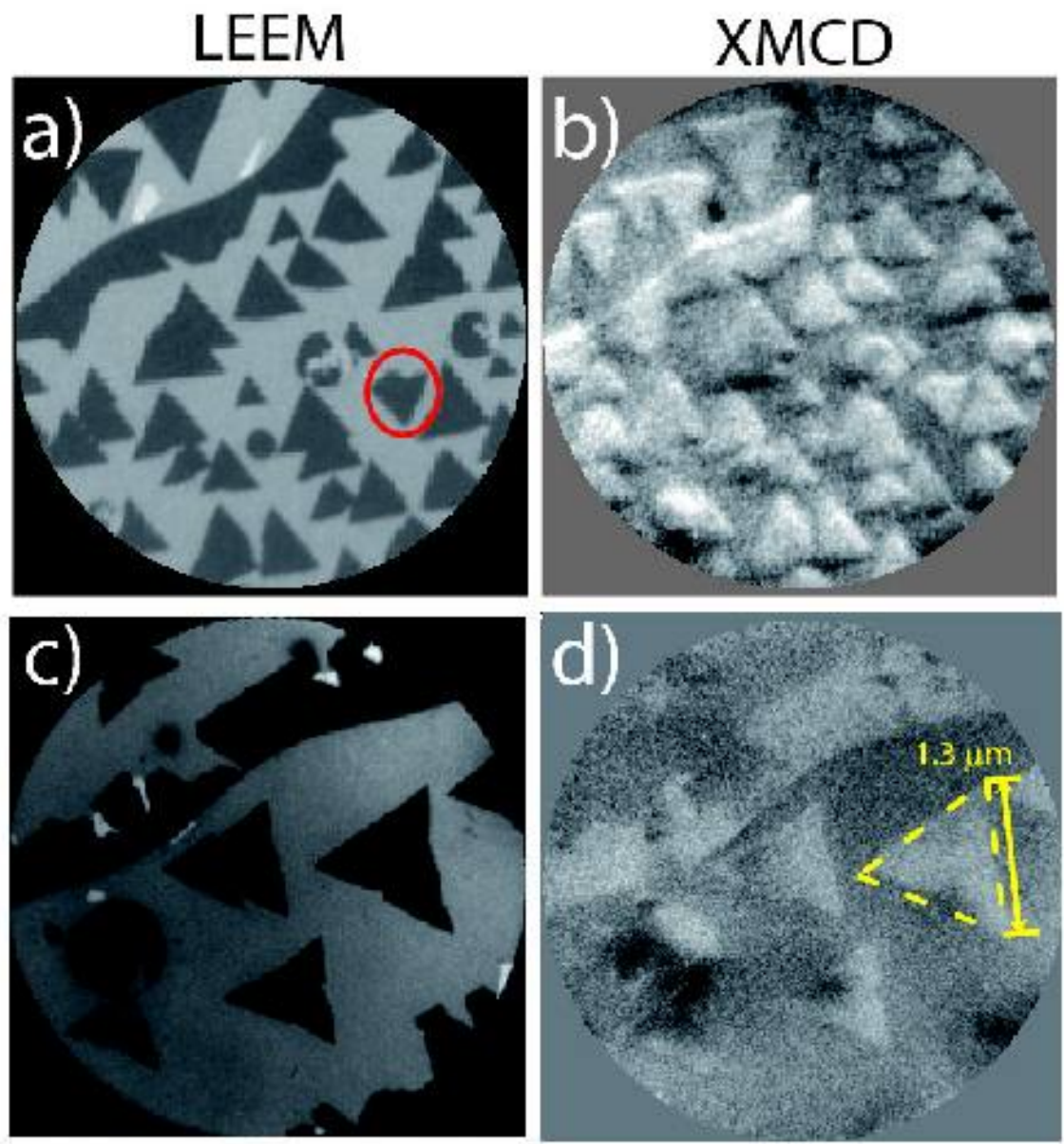

Fig. 1: a,c) LEEM images of 3 ML thick islands (dark triangles) on a 2 ML Co film (light grey

background) on top of Ru(0001) substrate. An island with different stacking order is indicated with a circle. b,d) Asymmetry (XMCD) images of the same regions as the LEEM images. Bright (dark) colors indicate the component of the magnetization of the islands along (opposite) the X-ray beam direction. All images have a field of view of 5 μm. The different island sizes in (a,b) versus (c,d) are due to different growth conditions (see text for details).

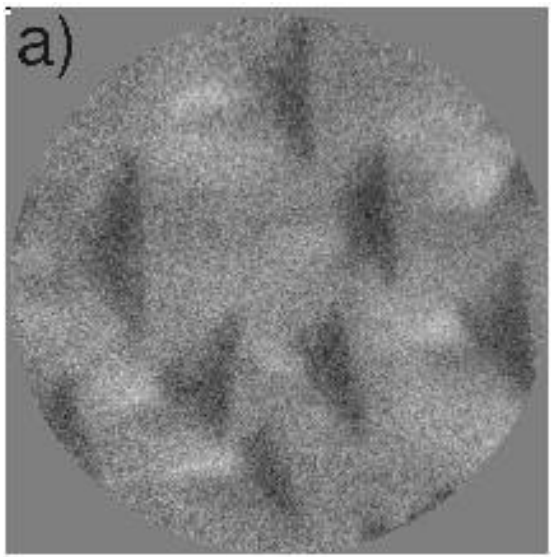
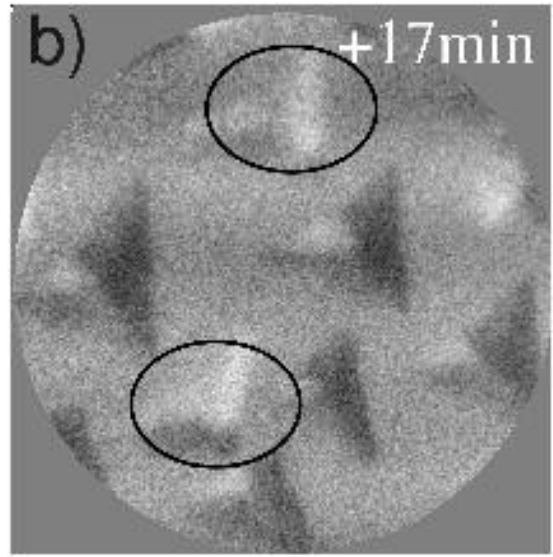
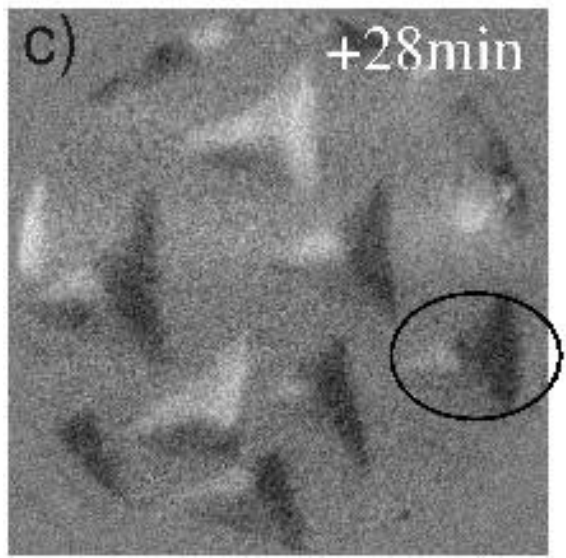
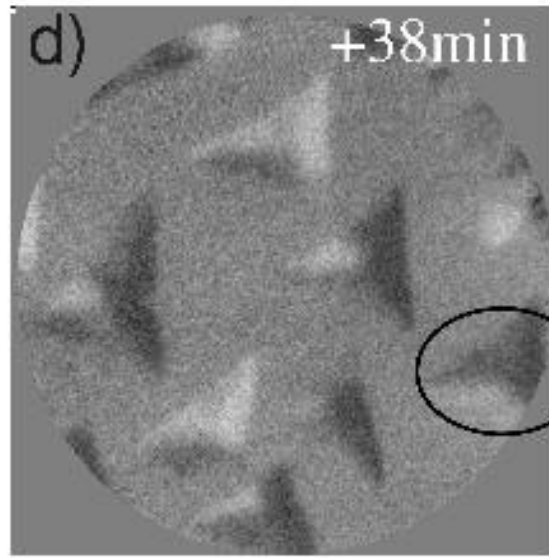

Figure 2: XMCD-PEEM images of the same film presented in Fig.1(c) (different region of the sample). A magnetic field pulse of 0.1 T has been applied before these images were acquired. The field of view is 5 μ m. Several consecutive images are shown, with the time after the first one indicated on each image. Note the chirality of the domain patterns and the changes on some of the islands with time, marked with circles.

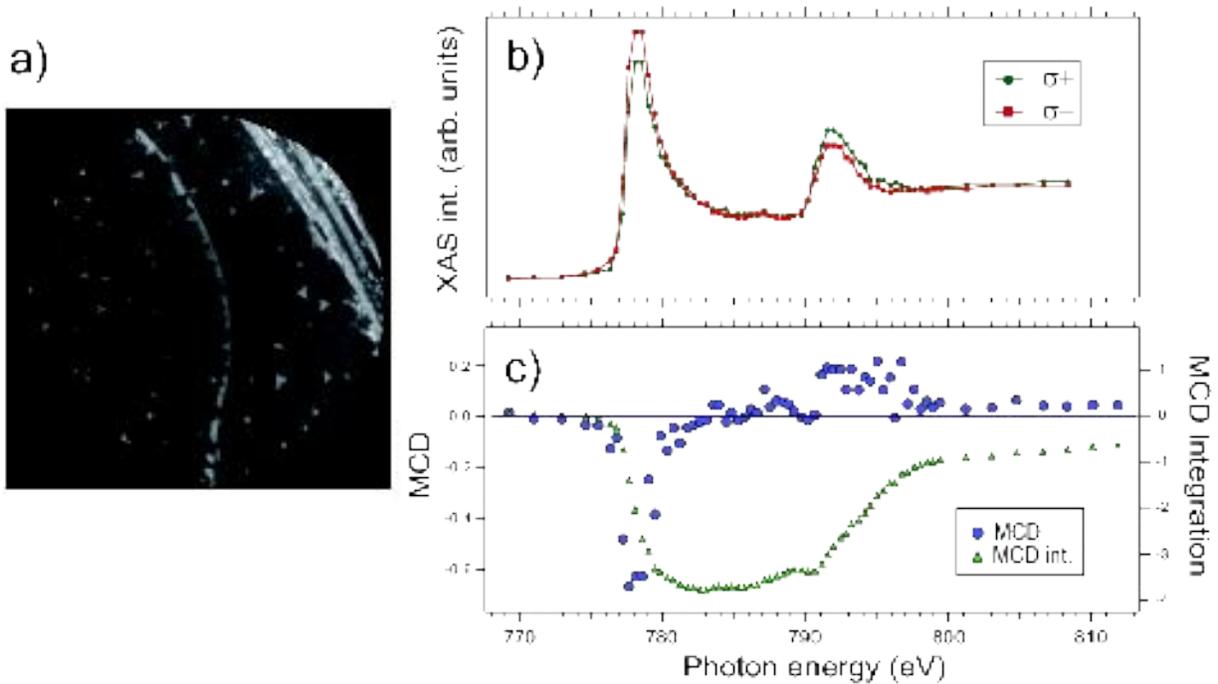

Figure 3: Dichroism on continuous films. a) LEEM image of a complete 3 ML film, with very small 4ML thick islands (light gray). The field of view is 10 μ m. b) XAS spectra of the film with circular polarization. c) MCD spectra of the 3 ML cobalt film (blue circles), and integrated MCD signal (green triangles).